\documentclass{vgtc}                          

\ifpdf
  \pdfoutput=1\relax                   
  \usepackage{graphicx}                
  \DeclareGraphicsExtensions{.pdf,.png,.jpg,.jpeg} 
\else
  \usepackage{graphicx}                
  \DeclareGraphicsExtensions{.eps}     
\fi%

\graphicspath{{figures/}{pictures/}{images/}{./}} 

\usepackage{microtype}                 
\PassOptionsToPackage{warn}{textcomp}  
\usepackage{textcomp}                  
\usepackage{mathptmx}                  
\usepackage{times}                     
\usepackage{cite}                      
\usepackage{tabu}                      
\usepackage{booktabs}                  
\usepackage{wrapfig}

\usepackage{amsfonts}
\usepackage{amsthm}
\usepackage{amsmath}
\usepackage{amssymb}
\usepackage{algorithm}
\usepackage{algorithmic}
\usepackage[shortlabels]{enumitem}
\usepackage{xcolor}
\usepackage{ulem}
\usepackage{comment}
\usepackage{xfrac}

\onlineid{5635}

\vgtccategory{Research}

\vgtcinsertpkg

\title{Neural Stream Functions}

\author{Skylar W. Wurster\thanks{e-mail: wurster.18@osu.edu}\\ %
        \scriptsize The Ohio State University %
\and Hanqi Guo\\ %
     \scriptsize The Ohio State University %
\and Tom Peterka\\ %
     \parbox{1.4in}{\scriptsize \centering Argonne National Laboratory} \\ %
\and Han-Wei Shen\\ %
     \scriptsize The Ohio State University}
     
\abstract{We present a neural network approach to compute stream functions, which are scalar functions with gradients orthogonal to a given vector field.
As a result, isosurfaces of the stream function extract stream surfaces, which can be visualized to analyze flow features.
Our approach takes a vector field as input and trains an implicit neural representation to learn a stream function for that vector field.
The network learns to map input coordinates to a stream function value by minimizing the inner product of the gradient of the neural network's output and the vector field. 
Since stream function solutions may not be unique, we give optional constraints for the network to learn particular stream functions of interest.
Specifically, we introduce regularizing loss functions that can optionally be used to generate stream function solutions whose stream surfaces follow the flow field's curvature, or that can learn a stream function that includes a stream surface passing through a seeding rake.  
We also discuss considerations for properly visualizing the trained implicit network and extracting artifact-free surfaces. 
We compare our results with other implicit solutions and present qualitative and quantitative results for several synthetic and simulated vector fields.%
} 

\CCScatlist{
  \CCScatTwelve{Flow Visualization}{Implicit Neural Representations}{Stream Surface Extraction}{}
}

\begin{document}


\firstsection{Introduction}

\maketitle


Flow data are pervasive across scientific fields, and visualizing these data give scientists insights into physical phenomena. 
Examples of areas of research that use flow data include direct numerical simulation (DNS) of the Navier-Stokes equations \cite{JHUTDB1} and climate and natural disaster research \cite{linn02_firetec1, linn_firetec2}.
One popular technique for visualization and analysis of these vector fields is \textit{stream surface} visualization.
Stream surfaces are surfaces tangential to the vector field, and therefore no flow goes through a stream surface.
Stream surfaces can also be used to analyze flow structure by segmenting the flow into distinct regions.

There are two families of approaches--explicit and implicit--for computing stream surfaces for a given vector field.
Originally proposed by Hultquist \cite{Hultquist92_streamsurface}, \textit{explicit methods} seed a set of points in the vector field, advect them using numerical integration with some interpolation scheme, and then mesh together the resulting streamlines to generate a stream surface.
Variations on this approach find seeds that maximize stream surface stretching \cite{barton17_stretchmaximizing} or perform distributed stream surface computation \cite{camp12_parallelstreamsurface, lu14_scalablestreamsurface}.
While intuitive and useful, these explicit approaches require some understanding about the flow data to guide placement of the seeding rake for the stream surface. 
Explicit approaches also have a high computational cost, since advection methods such as RK4 and meshing must be computed for many points, and must be re-computed for each surface of interest.

\textit{Implicit methods} for finding stream surfaces solve for a stream function such that isosurfaces of the scalar field are stream surfaces in the vector field.
As a consequence, a single implicit result yields a family of stream surfaces that can be extracted by varying the isovalue.
With many stream surfaces included in just one solution, features of the flow can be explored more quickly than explicit methods that require choosing seeding locations each time, since changing the isovalue and visualizing a new isosurface set is less expensive to compute than tracing particles and meshing a stream surface.
However, implicit methods have their limitations.
First, solving for an implicit solution can be computationally challenging and expensive.
Additionally, implicit approaches often do not allow control over seeding rakes, which may be useful for analyzing regions of interest.
Lastly, some implicit approaches may not provide a solution for all voxels within the domain, particularly in cases where the flow exits the domain \cite{stoter12_implicitintegralsurfaces} or is part of a recirculating current \cite{vanWijk93_implicitstreamsurface}.

Motivated by the limitations of current implicit solutions for stream surface extraction, we propose a novel approach to solve for a stream function of an arbitrary vector field.
Our approach uses an implicit neural network $f: D \rightarrow \mathbb{R}$ where $D$ is the domain and $f(x)$ is the stream function value at a point $x$.
Our network learns a stream function implicitly after being supplied only the vector field $V : \mathbb{R}^3
\rightarrow \mathbb{R}^3$ by minimizing the inner product of $\nabla_x f(x)$ and $V(x)$, where $\nabla_x$ is the spatial gradient, which means that isosurfaces of $f$ will lie tangent to the flow direction of $V$.
Our approach finds solutions efficiently and more accurately than those of the previous implicit approaches.
Our approach also allows optional control over the specific stream function solution via regularizing loss functions that can either (a) learn a stream function with stream surfaces oriented by flow curvature or (b) learn a stream function with a stream surface going through a seeding rake.
As an additional benefit, our compact neural representation has small storage and memory overhead for the solution, as opposed to a grid of the same size as the vector field.
Finally, we evaluate the sampling rate necessary to extract correct visualizations from the trained neural networks.
Though originally meshless, the networks must be sampled in order to visualize the scalar field result in modern visualization software at interactive speeds.

In summary, our contributions are threefold:
\begin{enumerate}
    \item An implicit neural representation of a stream function for a given vector field 
    \item Optional regularizing loss functions to generate \textcolor{black}{a solution} that provides a surface going through a seeding rake, or that creates stream surfaces oriented by the flow curvature
    \item A comprehensive evaluation against other state of the art techniques for implicit stream surface extraction
\end{enumerate}

\section{Related Works}

We survey related works for stream function visualization and implicit neural networks below.

\subsection{Stream Surface and Stream Function Visualization}

Stream functions and stream function visualization are a subset of flow visualization, which spans texture-based methods \cite{laramee04_textureflowvis}, geometry-based methods \cite{pauly09_geometricflowvis}, and feature identification and tracking methods \cite{post03_featureflowvis}.
For a comprehensive survey on stream surface visualization, we refer readers to Edmunds et al. \cite{edmunds12_streamsurfacesurvey}.
In our review, we categorize stream surface and stream function visualization into two categories: explicit and implicit methods.

Explicit stream surface visualization methods advect a seeding rake or sampled seeding curve in a vector field using numerical integration and generate a mesh from the resulting streamlines. 
Hultquist \cite{Hultquist92_streamsurface} first visualized stream surfaces and described an algorithm to add or remove seeding points along the front as it is advected to maintain even spacing within the rake. 
Ueng et al. \cite{Ueng96_efficientunstructuredstreamsurface} efficiently compute streamlines and streamsurfaces in unstructured grids. 
Rosanwo et al. \cite{Rosanwo09_dualstreamlineseeding} present an algorithm for ideal streamline seeding that improves domain coverage using an orthogonal vector field.
McLoughlin et al. \cite{McLoughlin09_quadstreamsurface} design a quad-based stream surface algorithm for efficient surface extraction.
Scheuermann et al. \cite{scheuermann01_tetrahedralstreamsurface} create a Barycentric coordinate stream surface calculation for tetrahedral grids. 
Van Gelder \cite{VanGelder01_curvilinearstreamsurface} creates an algorithm for stream surface generation in curvilinear grid flow data.
\textcolor{black}{Schneider} et al. \cite{schneider10_topologyawarestreamsurface} and Peikert and Sadlo \cite{Peikert09_topologicallyawarestreamsurface} create topologically aware stream surface algorithms.
Schafhitzel et al. \cite{Schafhitzel07_pointbasedstreamsurface} use a point-based algorithm for generating stream and path surfaces.
Schneider et al. \cite{Schneider09_4thorderstreamsurfaces} improve upon Hultquist's approach by using Hermitian interpolation to create smooth stream surfaces, giving better visualization results.
\textcolor{black}{Garth et al. \cite{garth08_integralsurfaces} generate accurate integral surfaces (both stream and path surfaces) that supports large time-varying vector fields with a data streaming implementation.}
Instead of creating stream surfaces that connect streamlines, Palmerius et al. \cite{Palmerius09_perpendicularstreamsurface} create surfaces that are orthogonal to the fluid flow at all points, and give a method to handle flow with non-zero helicity. 
\textcolor{black}{Similarly, Schulze et al. \cite{APAP_schulze12} generate surfaces in a vector field that are as-perpendicular-as-possible to the vector field.}
Zhang et al. \cite{Zhang21_surfriver} create a 2D exploration tool for analyzing stream surfaces traced in 3D.
Barton et al. \cite{barton15_stretchminimizing} keep the stream surface seed curve arc-length constant over advection in their stretch-minimizing approach for divergence-free flow.

Implicit methods for stream surface visualization extract a scalar valued stream function, and perform isosurface extraction to visualize stream surfaces. 
\textcolor{black}{Bhatia et al. \cite{bhatia13_HHD} provide a survey for the Helmholtz-Hodge Decomposition, which gives many examples of visualizing the scalar potential, which in our case is called the stream function.}
Van Wijk \cite{vanWijk93_implicitstreamsurface} was the first to explore implicit extraction of stream surfaces, in which he defines two computational methods to generate the 3D scalar field for isosurface extraction: one using the convection equation and the other using an algorithm that backward traces ink from domain boundaries.
Kenwright and Mallison \cite{kenwrightmallison92_dualstreamfunctions} use the dual stream function approach to illustrate an $f$-$g$ graph.
Cai and Heng \cite{Cai97_principalstreamsurface} iteratively build the 3D scalar field for isosurface extraction using a Taylor expansion of the scalar potential for the normal direction to the velocity field such that the resulting stream surfaces lie on the recitifying plane of the Frenet frame, but the approach is limited to irrotational flow. 
Westermann et al. use van Wijk's ink tracing approach in a multiscale method for time surface visualization of flow features, which have constant isovalue over multiple timesteps. 
Beale \cite{Beale97_dualstreamfunction} solves for the dual stream function approach using a finite volume method, and finds streamlines as intersections in the two scalar fields.
Reztsov and Mallison \cite{reztsovmallison98_swirlingflows} use the dual stream function approach for 3D swirling flow on solenoidal vector fields, and Li and Mallison \cite{limallison06_axissymmetry} extend Reztsov and Mallison's approach to flow with axissymmetry on rectilinear grids.
It has been proven that the dual stream function approach cannot accurately represent the vector field near critical points in the vorticity field \cite{Graham2000_clebschnearcriticalpoints}.

Stöter et al. \cite{stoter12_implicitintegralsurfaces} create a method for generating implicit integral surfaces by advecting two seeding scalar fields over time using a computed flow map for the vector field and taking intersections of two 3D isovolumes.
Since this method requires both seeding as well as streamline advection, the approach is analogous to precomputing many explicitly calculated surfaces, caching them, and retrieving them on demand with an isovolume intersection.
Instead of supplying two scalar seeding functions and performing numerical integration for advection as Stöter et al. do, we instead learn a single scalar stream function from which isosurfaces directly visualize stream surfaces, with no advection or seeding required.
Unlike our approach, none of the above methods model a stream function using an implicit neural representation, which captures the vector field in a compact form.

\subsection{Implicit Neural Representations}

Implicit neural representations (INRs) are neural networks that represent a signal, such as an image, audio, or 3D scalar field, continuously in the input domain.
For instance, a network $f: \mathbb{R}^3 \rightarrow R$ maps coordinate $x$ to the scalar value at $x$ in some scientific scalar field. 
They are generally fully connected (non-convolutional) neural networks.
Once a network is trained to represent a signal, the signal can be reconstructed by sampling the network at desired locations.
INRs have shown to fit signals such as images, signed distance fields, sound, and video \cite{sitzmann19_siren}, as well as scientific data such as scalar fields \cite{Lu21_compressiveneuralrep}.
One benefit of INR is that the representation is continuous.
Any point can be queried in the input domain, making this a meshless representation for the signal.
Another benefit is the inherent compressive capability since the original signal is represented with a neural network that generally requires less memory to store than the signal sampled on a regular grid. 

A sinusoidal representation network (SIREN) \cite{sitzmann19_siren} is an architecture for INR that uses sinusoidal activations in a fully connected neural network to model images, signed distance fields, and audio, as well as their derivatives, more accurately than other activation functions.
Lu et al. \cite{Lu21_compressiveneuralrep} use a modified SIREN network to model a scalar field.
They add residual connections and quantize the network weights after training for additional storage reduction.
Lindel et al. \cite{lindell21_autoint} efficiently perform neural volume rendering by pre-computing a SIREN network's gradient with respect to the input, and training this gradient network.
Since the gradient and integral networks share the same weights, after training, the network has learned the integral of the trained function. 
They use this integral network with dynamic ray stepping to improve neural volume rendering speed over typical Monte-Carlo methods.
Sitzmann et al. use an INR of light fields to speed up neural rendering speeds \cite{sitzmann2021_lfns}.
Mildenhall et al. create neural radiance fields \cite{mildenhall2020_nerf}, which use an implicit neural network to synthesize new views for a scene given a set of images from the scene.
Our approach uses an INR for a stream function representation, and we train by minimizing the gradient of the neural network and the supplied vector field.
Unlike most other INR applications, our approach is learning an unknown signal, regularize only by the network's gradient.

\section{Background}

\label{background}
In this background section we will explain how stream functions are mathematically related to the vector field, the definition of the Frenet frame for a 3D vector field, and how gradients of neural network outputs \textcolor{black}{may} be used to trained the network.

\subsubsection{Stream Functions}
\label{backgroundstreamfunctions}
In 3D, a stream function $f$ for a vector field $V$ is defined such that $\nabla f \cdot V = 0$, where $f$ is a scalar function and $\nabla$ is the gradient operator.
As a result, isovalues in $f$ create stream surfaces for $V$. 
Stream surfaces are surfaces that are everywhere tangent to the vector field, just as stream lines are everywhere tangent to the vector field.

\subsubsection{Frenet Frame}
\label{backgroundfrenetframe}

\begin{wrapfigure}{r}{0.2\textwidth}
    \vspace{-0.25in}
    \includegraphics[width=3.5cm]{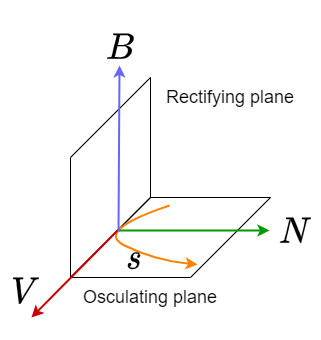}
  \caption{Depiction of the normal ($N$) and binormal ($B$) directions in the Frenet frame \cite{Cai97_principalstreamsurface}. An example of a streamline is drawn as $s$.}
\label{fig:frenet_frame}
\end{wrapfigure}

In the Frenet frame (also known as the $TNB$ frame) in \autoref{fig:frenet_frame}, there are two principal directions for a space curve $s(a(t))$, where $a(t)$ is the arclength parameterized by $t$.
We assume the curve is regular such that $a'(t) \neq 0$.
At a specific point on the curve, there is a tangent direction labeled $T$, a normal direction labeled $N$, and a binormal direction labeled $B$.
The tangent direction is the derivative 
\begin{equation}
    T = \frac{\partial s}{\partial a}.
\end{equation}

We note that $T \cdot T = 1$, and take the derivative with respect to arclength, giving $T \cdot \frac{\partial T}{\partial a} = 0$.
Therefore, the derivative $\frac{\partial T}{\partial a}$ must be orthogonal to the tangent direction $T$.
The normal direction in the Frenet frame $N$ is defined $\kappa N = \frac{\partial T}{\partial a}$, where $\kappa$ is a scalar representing the strength of the curvature at the point.
Finally, the binormal direction is defined by $B = T \times N$, and is orthogonal to both vectors $T$ and $N$.

The vector fields we use in this paper already represent the first derivative of space curves, i.e., $V = (u,v,w) = \frac{\partial s}{\partial t}$, and the original space curves are not saved.
Therefore, the calculations for the normal and binormal are given in other terms that do not rely on the space curve $s$.
The binormal direction, $B$, is calculated by 
\begin{equation}
    \label{binormal_eq}
    B = J_V V \times V,
\end{equation}
where $J_V$ is the Jacobian of the vector field $V$.
The normal direction is then calculated as 
\begin{equation}
    \label{normal_eq}
    N = B \times V.
\end{equation}
Together, the $N$ and $B$ vector fields span the space of all orthogonal vector fields for $V$.

\setlength{\columnsep}{6pt}
\setlength{\intextsep}{2pt}

As shown by Cai and Heng \cite{Cai97_principalstreamsurface}, the rectifying plane (defined as the plane with a normal direction $N$), gives stream surface visualizations that show flow curvature. 
In their approach, they solve for the stream function $f$ such that $\nabla f = N$.
Since $N \cdot V = 0$ by definition and $\nabla f \cdot V = 0$, $f$ is a stream function.
The resulting stream surfaces of this construction align with the local curvature of flow.

In Cai and Heng's approach, the flow data must be irrotational, meaning the vector fields have an exact scalar potential.

\subsubsection{Neural Network Gradient Training}
\label{backgroundneuralnetworkgradienttraining}

\begin{figure}[ht]
\centering
  \includegraphics[width=3in]{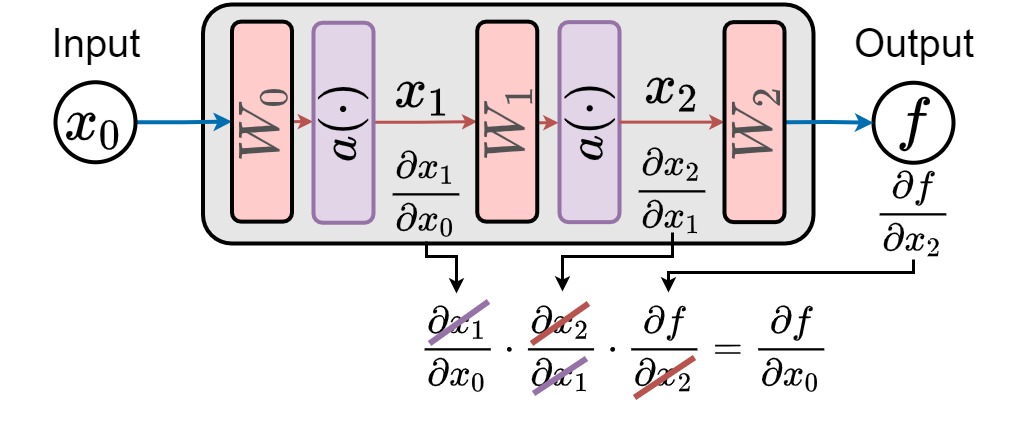}
  \caption{An example of how gradients of the network output can be taken with respect to the network input. Input $x_0$ is fed through linear layers, depicted by $W_i$, where is $W_i$ is a weight matrix, followed by a nonlinear activation function $a(\cdot)$.}
  \label{fig:chain_rule}
\end{figure}

The chain rule can be used to compute closed form neural network gradients, and is used to train neural networks when computing the derivative of the loss function with respect to the network's weights.
Alternatively, the chain rule can also be used to generate derivatives of the output with respect to the input.
An example is shown in \autoref{fig:chain_rule} with a two layer fully connected network with a single input $x$ and single output $f$. 
Derivatives of each layer's output with respect to its input can be computed locally because the derivative of the layer's operation is known. 
For instance, assume $\sin(\cdot)$ is used as the activation function $a(\cdot)$ in the network in \autoref{fig:chain_rule}.
Then we have

\begin{equation}
    \begin{split}
        x_1 =  \sin(x_0 W_0), ~~ x_2 = \sin(x_1 W_1), ~~ f = x_2 W_2 \\
        \frac{\partial x_1}{\partial x_0} =  W_0 \odot \cos(x_0 W_0) \\
        \frac{\partial x_2}{\partial x_1} = W_1 \cos(x_1 W_1)^T \\
        \frac{\partial f}{\partial x_2} = W_2,
    \end{split}
\end{equation}
where $\odot$ is the element-wise multiplication and each layer of weights.
Then we can directly calculate $\frac{\partial f}{\partial x_0}$ with

\begin{equation}
    \label{eq:chain_rule}
    \begin{split}
        \frac{\partial f}{\partial x_0} & = 
        \frac{\partial x_1}{\partial x_0} \cdot  \frac{\partial x_2}{\partial x_1} \cdot \frac{\partial f}{\partial x_2} \\
        & = \left(W_0 \odot \cos(x_0 W_0)\right) \left(W_1 \cos(x_1 W_1)^T\right) W_2.
    \end{split}
\end{equation}
The idea extends to multiple inputs and outputs.

This is similar to work that uses neural networks to solve ordinary and partial differential equations \cite{berg18_nnpde, sirgnano18_dgm, raissi19_physics, raissi17_physicsI, raissi17_physicsII}.
This usage of the chain rule has been explored in other scientific visualization works for sensitivity analysis of trained neural networks \cite{hazarika20_NNVA, he20_insitunet, shi22_gnnsurro}.
Instead of using the chain rule for post-hoc analysis of the trained model, we use the chain rule to calculate the gradients for use during network training.

\section{Approach}

\begin{figure*}[ht]
\centering
  \includegraphics[width=\textwidth]{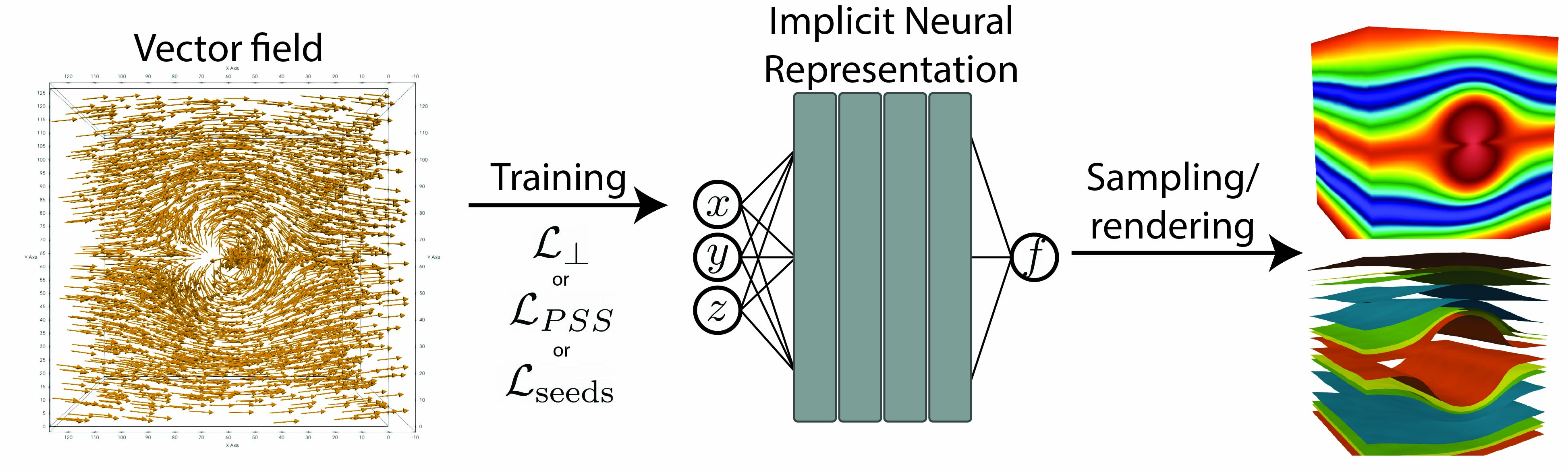}
  \caption{An illustration of our approach which solves for a stream function. Given a vector field, an implicit neural representation (INR) is trained with optional regularization to generate a stream function with specific characteristics. Once the network is trained, the scalar field stream function can be sampled on a grid and visualized in software such as ParaView, or the network can be queried directly to perform direct neural rendering to visualize stream surfaces.}
  \label{fig:pipeline}
\end{figure*}

Our approach uses a fully connected neural network $f$ to model a stream function for a vector field $V$, where the network's input is a spatial coordinate and the network's output is the stream function value at the input coordinate.
In \autoref{neuralstreamfunction}, we discuss how we train a neural network to learn a stream function for a vector field without a ground truth stream function to learn from directly.
In \autoref{frenetframe}, we discuss how we train our network to learn a stream function that generates surfaces relevant to the flow curvature such as those by Cai and Heng \cite{Cai97_principalstreamsurface}.
In \autoref{implicitseedingpoints}, we explain how our model can be trained to generate a stream function with a stream surface that goes through a seeding rake, making implicit stream surfaces more like those in explicit methods.

\subsection{Neural Stream Function}
\label{neuralstreamfunction}

Given only a vector field $V$, we generate one implicit stream function solution (of the infinitely many) for $V$, such that isovalues of the stream function create stream surfaces.
We model the stream function as a neural network $f(x): \mathbb{R}^3 \rightarrow \mathbb{R}$ where $f(x)$ is the stream function value at coordinate $x$.
Recall that the stream function for the supplied vector field $V$ is not known in advance.
We rely on two facts outlined in \autoref{background}: (1) a stream function solution will have a gradient that is everywhere perpendicular to the vector field, and (2) the chain rule can be used to create a closed form solution to the gradient of a neural network, which can then be used in a loss function to supervise training. 
A neural network $f$ can learn a stream function by training the network's gradient to be orthogonal to $V$ by minimizing the following loss function:

\begin{equation}
    \mathcal{L}_{\perp} = \frac{1}{|P|}\sum_{x \in P} |\nabla_x f(x) \cdot V(x)|,
\end{equation}
where $P$ is a set of sampled coordinates.

To train a network using this loss function, the vector field's domain is normalized to $[-1, 1]^3$, with the sampled vector field values evenly spaced in the domain.
During each iteration of training the network, \textcolor{black}{$b$ (the batch size hyperparamter)} location-vector pairs are extracted to sets ($P$, $W$) respectively, where $P$ is a grid point's location, and $W$ is the vector field value at that point. 
The sampled locations $P$ are fed to the neural network as a batch, and the output values are then differentiated through the network with respect to the input locations.
This gradient results in $\nabla_x f(x)$ for each $x\in P$.
The dot product of these gradients with the vector field values $W$ at the sampled locations is taken, and averaged in the final loss value.

Since our loss function involves a quantity that is the result of a gradient taken through the neural network, we use the SIREN architecture \cite{sitzmann19_siren} as the model to learn the stream functions, with residual connections added by Lu et al. \cite{Lu21_compressiveneuralrep}.
As shown by the Sitzmann et al., the SIREN architecture can learn signals by training with the gradient alone, and can do so with higher accuracy than other implicit representations that use activations such as ReLU or sigmoid.

\begin{figure*}[ht]
\centering
  \includegraphics[width=0.9\textwidth]{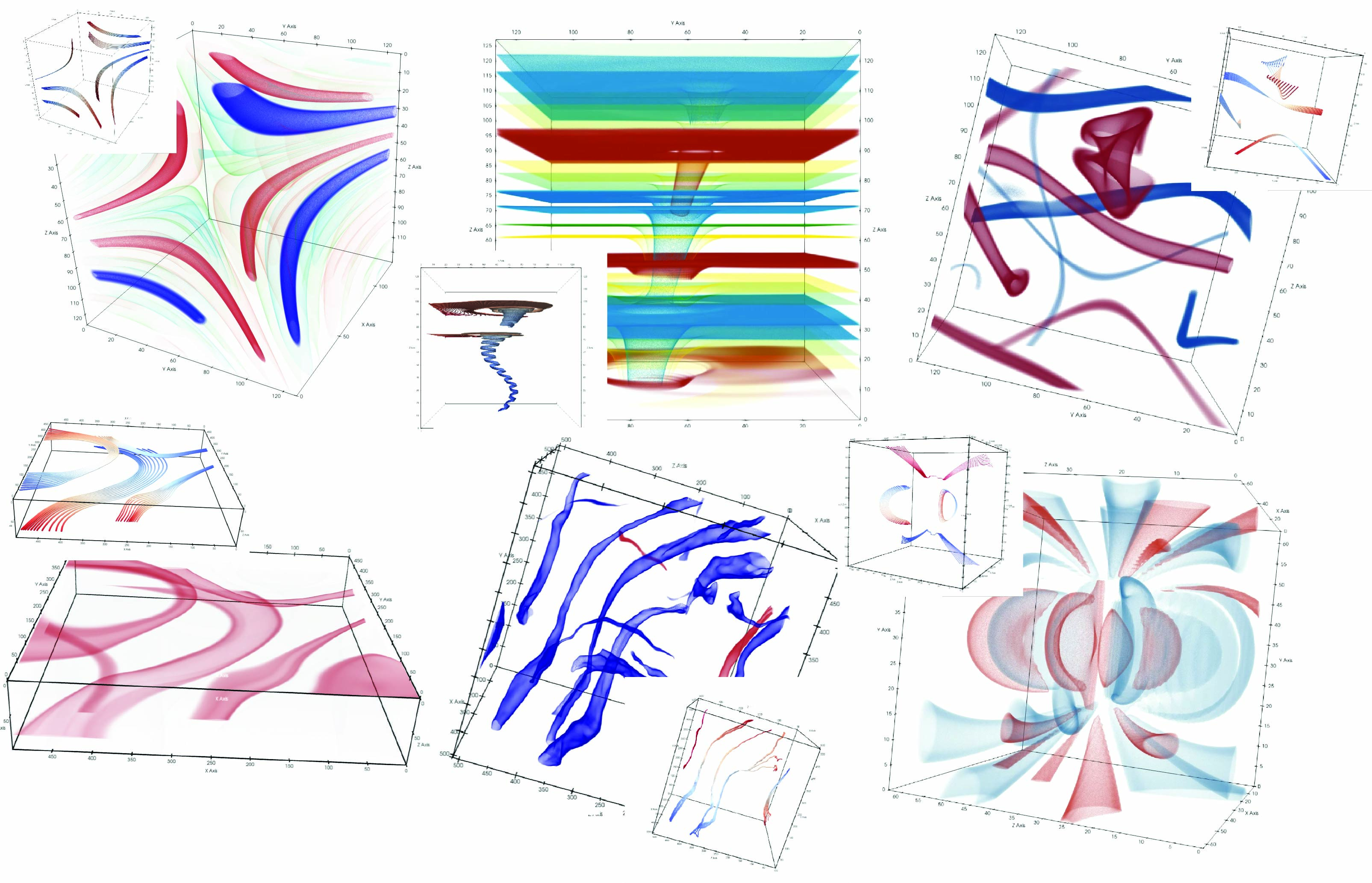}
  \caption{Stream functions created with our method rendered using spiking transfer functions or isosurface extraction. Streamlines are visualized as black tubes. Renders of streamlines generated from multiple rakes are added on each render for comparison between our stream surfaces and the explicitly calculated streamlines, colored by integration time. Datasets from top left to bottom right are vortices, tornado, ABC, Isabel, isotropic, and classic hill.}
  \label{fig:streamfunctions}
\end{figure*}

\subsection{Stream Surfaces Following Flow Curvature}
\label{frenetframe}

Since there may be infinite solutions for a stream function $f$ such that $\nabla f \cdot V = 0$, the isosurfaces in the solution may not depict stream surfaces of interest.
One approach by Cai and Heng \cite{Cai97_principalstreamsurface} to provide more useful stream surfaces uses the Frenet frame, described in \autoref{backgroundfrenetframe}.
To incorporate this into our method, we train our neural network such that the gradient of the network is parallel to the principal normal direction in the Frenet frame.
Specifically, we use the loss function
\begin{equation}
    \mathcal{L}_{PSS} = \frac{1}{|P|}\sum_{x \in P} 1 - \left(\frac{\nabla f(x) \cdot N(x)}{||\nabla f(x)||_2 * ||N(x)||_2}\right)^2
\end{equation}
where $x$ is a sample coordinate from a set of points $P$, $N$ is the principal normal direction and $||\cdot||_2$ is the L2 norm.
With this loss function, the surfaces generated will follow flow curvature, and may depict the flow structure more clearly.

We pre-process the vector field to find its principal normal direction $N$ via a first order central difference estimation of the gradients to calculate the vector field's Jacobian, and then use \autoref{binormal_eq} and \autoref{normal_eq} to calculate $N$.
This loss function is used in lieu of the standard loss function $\mathcal{L}_{\perp}$, as it would be redundant to minimize both quantities.

\subsection{Implicit Seeding Rake}
\label{implicitseedingpoints}

One limitation of many implicit approaches is that it may be difficult to extract stream surfaces of interest that pass through some seeding rake.
In our approach, we propose a solution to this which allows a more traditional explicit seeding rake choice to guide the solution stream function to have a surface going through the rake given.
Specifically, we use the loss function 

\begin{equation}
     \mathcal{L}_{\text{seeds}} = \frac{1}{|S|}\sum_{s \in S} f(s)
\end{equation}
where $f$ is the neural network, and $s$ is a seed position within the set of seeds $S$ densely sampled from the rake desired.
In other words, the loss term is encouraging the network to learn $f(s \in S) = 0$ such that after training, the isosurface for isovalue 0 should go through the seeding rake.
The isovalue chosen of 0 may be arbitrary, as the final layer of the neural network is a linear combination of previous node values plus a bias term, and so the bias term can offset the isovalue to be whatever needed. 
Each iteration during training, the points from the seeding rake $S$ are fed through the neural network and the output is averaged to calculate this loss value, which is added to the standard loss $\mathcal{L}_{\perp}$.
\textcolor{black}{In practice, this loss is added to either $\mathcal{L}_{\perp}$ or $\mathcal{L}_{PSS}$ as is defined in \autoref{neuralstreamfunction} and \autoref{frenetframe}, respectively, such that $\mathcal{L}_{\perp}$ or $\mathcal{L}_{PSS}$ is responsible for making the scalar field gradient orthogonal to the flow, and $\mathcal{L}_{\text{seeds}}$ is responsible for ensuring a stream surface goes through the seeding rake.}

\section{Evaluation}

We evaluate our approach across many datasets for both qualitative and quantitative measures. 
Our code is implemented in Python 3.9 using PyTorch \cite{NEURIPS2019_pytorch} for the deep learning framework. 
All models use single precision floating point arithmetic and are trained on a single NVidia A100-SXM4-40GB. 
All code is available at \url{https://github.com/skywolf829/NeuralStreamFunction}.

Our evaluations are both qualitative and quantitative.
We compare both the quality of the visualized isosurfaces in the rendered images and orthogonality error of the stream function $f$.
We measure the error at each voxel with the metric:
\begin{equation}
\label{errorquant}
    Err_{\perp}(x) = \left|\frac{\pi}{2} - \cos^{-1}\left(\frac{\nabla_x f(x) \cdot V(x)}{||\nabla_x f(x)||_2 \cdot ||V(x)||_2}\right)\right|,
\end{equation}
where $x$ is the coordinate, $f$ is the learned stream function, and $V$ is the vector field.
The metric measures the angle error (in radians) from perpendicular of the stream function gradient with respect to the vector field at a particular coordinate.
Using this calculation on each voxel allows us to visualize the error volume to identify regions of high error, and it also allows us to list statistics about the error for various methods in our evaluation.
When evaluating this error using a trained neural network, we directly evaluate $\nabla_x f(x)$ from the network using the chain rule as discussed in \autoref{backgroundneuralnetworkgradienttraining}, and do not rely on finite difference methods.
However, the two methods with which we compare generate discrete scalar fields, and as such, the first order gradient has to be approximated using a finite difference method in order to evaluate $\nabla_x f(x)$ to compute $Err_{\perp}$.

In \autoref{datasets}, we introduce the vector field datasets for our experiments.
In \autoref{training}, we discuss how our network is trained and the architecture for the networks we use.
In \autoref{evaluation}, we analyze error and visualize results from each of the three loss functions and compare with related works where applicable, and we discuss which loss function to use in which situation.
Lastly, we discuss sampling rates necessary to accurately depict surfaces learned in \autoref{neuralrendering}.

\begin{table}[h]
\centering
\caption{Dataset names and their sizes.}
\begin{tabular}{l|c}
Dataset  & Volume size \\ \hline \hline
Flow past cylinder & $128^3$     \\
Vortices    & $128^3$    \\
Tornado & $128^3$ \\
ABC flow & $128^3$     \\
Isabel & $500^2\times100$ \\
Isotropic & $512^3$  \\
Classic Hill & $64^3$ \\
Delta 40 & $128\times 41^2$ 

\end{tabular}
\label{datasets_table}
\end{table}

\subsection{Datasets and Preprocessing}
\label{datasets}

We use eight datasets \textcolor{black}{(seven incompressible, one compressible}) for evaluation of our approach, with their names and sizes listed in \autoref{datasets_table}.
Flow past cylinder and vortices are two analytically defined datasets which are given by Cai and Heng\cite{Cai97_principalstreamsurface}.
We sample at $128^3$ within the same domain as Cai and Heng for an accurate comparison.
Tornado is an analytically defined dataset provided by Crawfis \cite{Crawfis03_tornado}.
\textcolor{black}{We note that due to a conditional statement in the code to generate the tornado data, the result is not incompressible (non-zero divergence) everywhere.}
ABC flow is Arnold–Beltrami–Childress flow, which is an analytically defined vector field known for having potentially chaotic streamline trajectories.
Isabel is a single timestep of the hurricane Isabel dataset produced by the Weather Research and Forecast (WRF) model, created by of NCAR and the U.S. National Science Foundation (NSF)\textcolor{black}{, and is the only compressible dataset we test with.}
Isotropic is the result of a 3D DNS for isotropic flow with Reynolds number $R_{\lambda}\sim 433$ , created and hosted by John's Hopkins Turbulence Database \cite{JHUTDB1}.
Classic hill and delta 40 are two datasets defined and supplied by Chern et al. \cite{chern17_insidefluids}, which are an analytical dataset and a result from a DNS of airflow over a wing, respectively.

\subsection{Network Architecture and Training}
\label{training}
Our networks follow the architecture presented in neurocomp \cite{Lu21_compressiveneuralrep}, which is a SIREN \cite{sitzmann19_siren} architecture with residual connections. 
\textcolor{black}{A SIREN architecture uses only fully connected layers of the same width with a sinusoidal activation function after each layer except the last. The residual connections were introduced to the SIREN architecture for scientific data by Lu et al. \cite{Lu21_compressiveneuralrep}, and connect the output of one layer with the output of a layer deeper in the network by adding the values.}
\textcolor{black}{Experimentally, the larger the model, the smaller the error. At the same time, larger models take longer to train. We find that a model of size 4 layers deep and 512 neurons wide is a good middle ground.}
We experimentally find that a batch size of between 0.1\% and 1\% of the total number of voxels \textcolor{black}{in the vector field being trained with} gives a good quality-to-training speed ratio\textcolor{black}{, where the sampled batch is pairs of $(x,y,z)$ locations (with extents $[-1, 1]^3$) and that location's vector value $(u,v,w)$.}
Only voxel grid points are used for training the network; no interpolation between grid points is used when training the network. 
\textcolor{black}{Increasing the number of points sampled per iteration can improve the resulting accuracy at the cost of larger training times and VRAM usage.}

\textcolor{black}{Each iteration, the network's output is queried for the batch of $(x,y,z)$ locations, and the loss function of choice is applied on the output of the network and the vector field's $(u,v,w)$ values for each location.}
We use a learning rate of $5 \times 10e -5$ with the Adam optimizer \cite{kingma15_adam}.
We decay the learning rate by a factor of 10 every 3,333 iterations, and train for a total of 10,000 iterations. 

\begin{table}[h]
\caption{Network training data for models evaluated. GPU memory reported is maximum GPU memory reserved during network training. All networks are 4 layers of 512 neurons, and cost 4,118 KB to save to storage.}
\begin{tabular}{l|ccc}
Dataset      & Samples per iter. & Train time & GPU memory \\ \hline
Cylinder     & 10,000            & 2m 18s        & 0.549 GB   \\
Vortices     & 10,000            & 2m 19s        & 0.549 GB   \\
Tornado      & 10,000            & 2m 20s        & 0.549 GB   \\
ABC flow     & 10,000            & 2m 19s        & 0.549 GB   \\
Isabel       & 125,000           & 26m 2s        & 7.017 GB   \\
Isotropic    & 100,000           & 37m 16s       & 9.401 GB   \\
Classic Hill & 10,000            & 2m 14s        & 0.547 GB   \\
Delta 40     & 10,000            & 2m 14s        & 0.546 GB  
\end{tabular}
\label{trainingtable}
\end{table}

Training settings and computation resources for networks that learn stream functions for each dataset are displayed in \autoref{trainingtable}.
Training speeds depend on (1) the size of the model and (2) the number of points sampled per iteration, whereas GPU memory additionally depends on the size of the vector field, since the vector field is stored in GPU memory for efficient access to values during the loss function computation.
For the datasets of size $128^3$ or smaller, the network takes under 2.5 minutes to train, uses roughly 0.5GB of GPU memory during training, and the model occupies just over 4MB in storage.
Note that during inference, the model uses less GPU memory because network gradients do not need to be saved as they do during training.
We find that 4 layers of 512 neurons is a sufficient size network to achieve low error with acceptable training times, though larger networks may increase accuracy at the cost of longer training times and larger memory requirements.

\begin{table}[h]
\caption{Error metric $Err_{\perp}$ for the learned stream functions (in degrees). Each result listed is the median/mean error over the entire volume. 
Empty entries are used where the setting is not ran.}
\centering
\begin{tabular}{l|cccc}
Dataset      & $\mathcal{L}_{\perp}$ & $\mathcal{L}_{PSS}$ & $\mathcal{L}_{\perp} + \mathcal{L}_{\text{seeds}}$ & Vorticity $\mathcal{L}_{\perp}$ \\ \hline
Cylinder     & \sfrac{0.009}{0.017}                       & \sfrac{0.779}{1.162}      & --                 & --           \\
Vortices     & \sfrac{0.013}{0.029}                       & \sfrac{3.205}{4.404}      & --                 & --           \\
Tornado      & \sfrac{0.020}{0.060}                       & \sfrac{3.816}{5.311}      & --                 & \sfrac{0.052}{0.286}  \\
ABC flow     & \sfrac{0.020}{0.053}                       & --       & \sfrac{0.043}{0.131}        & --           \\
Isabel       & \sfrac{0.038}{0.119}                       & --               & \sfrac{0.088}{0.238}        & --           \\
Isotropic    & \sfrac{1.557}{3.050}                       & --               & --                 & --           \\
Classic Hill & \sfrac{0.020}{0.092}                       & --               & --                 & \sfrac{0.028}{0.146}  \\
Delta 40     & \sfrac{0.009}{0.030}                       & \sfrac{1.291}{1.995}      & \sfrac{0.018}{0.072}        & \sfrac{0.073}{0.543}                                    
\end{tabular}
\label{quanterrortable}
\end{table}

\subsection{Neural Stream Surface Evaluation}
\label{evaluation} 

In our approach, we design three different loss functions that result in stream functions tailored to learn stream surfaces of interest.
Each loss function has different use cases and related works to evaluate.
We evaluate our $\mathcal{L}_{\perp}$ loss function in \autoref{baselineeval}, and show that it can be used to extract \textit{vortex tubes} when training on a vorticity field, and compare our approach with another implicit representation \cite{chern17_insidefluids} that can extract vortex tubes.
We evaluate our $\mathcal{L}_{PSS}$ loss function and compare with related work by Cai and Heng \cite{Cai97_principalstreamsurface} in \autoref{PSSevaluation}.
We evaluate our seeding rake loss function $\mathcal{L}_{\text{seeds}}$ in \autoref{seedseval}, and show that it creates stream functions that extract a surface through the seeding rake.
Lastly, we comprehensively compare our three loss functions and discuss which loss function to use in \autoref{discussion}.

\subsubsection{$\mathcal{L}_{\perp}$ Evaluation}
\label{baselineeval}
We present quantitative results for stream functions learned using $\mathcal{L}_{\perp}$ for the datasets listed in the second column of \autoref{quanterrortable}, which uses the median orthogonality error, defined in \autoref{errorquant}.
With no other fully implicit approaches to compare, we evaluate our approach quantitatively with $Err_{\perp}$ and qualitatively with a visualization of learned stream functions in \autoref{fig:streamfunctions}.
Our quantitative results for a majority of datasets show errors that are two orders of magnitude smaller than 1 degree throughout the entire domain and across any of the possible stream surface extracted, since the median $Err_{\perp}$ is calculated globally, and different stream surfaces can be extracted with different isovalues.
Therefore, the solution stream functions from our approach do not only create one stream surface with low error, but an entire family of stream surfaces with low error that can be visualized via isosurface extraction.

Visualizations of the results from our approach using $\mathcal{L}_{\perp}$ are shown in \autoref{fig:streamfunctions} using isosurface extraction or volume rendering with spiking transfer functions.
We verify our accuracy by tracing streamlines seeded on an extracted isosurface, or within an extracted stream tube in \autoref{fig:streamfunctions} as black lines.
We expect these streamlines to lie on the stream surface extracted or to stay within a stream tube.
The streamlines traced validate our low error calculation since the streamlines do not pass through stream surfaces/tubes extracted, and follow the shape of the stream surface extracted.

\textbf{Vortex Tubes}. As described, our approach applied to a velocity field will create a scalar field where isosurfaces represent stream surfaces.
However, our approach (with no modifications) applied to a vorticity field will create a scalar field where isosurfaces represent \textit{vortex tubes}, which are surfaces that integral curves traced in the vorticity field do not pass through.
A work by Chern et al. \cite{chern17_insidefluids} generates an implicit solution for vortex tubes as a by-product of their approach, which models a vector field with a spherical Clebsch representation.
We compare results from their method with results from our method by taking the curl of the vector field using a first-order central difference method for vector field gradient estimation and then extracting vortex tubes using each method.

\begin{figure}[ht]
\centering
  \includegraphics[width=\linewidth]{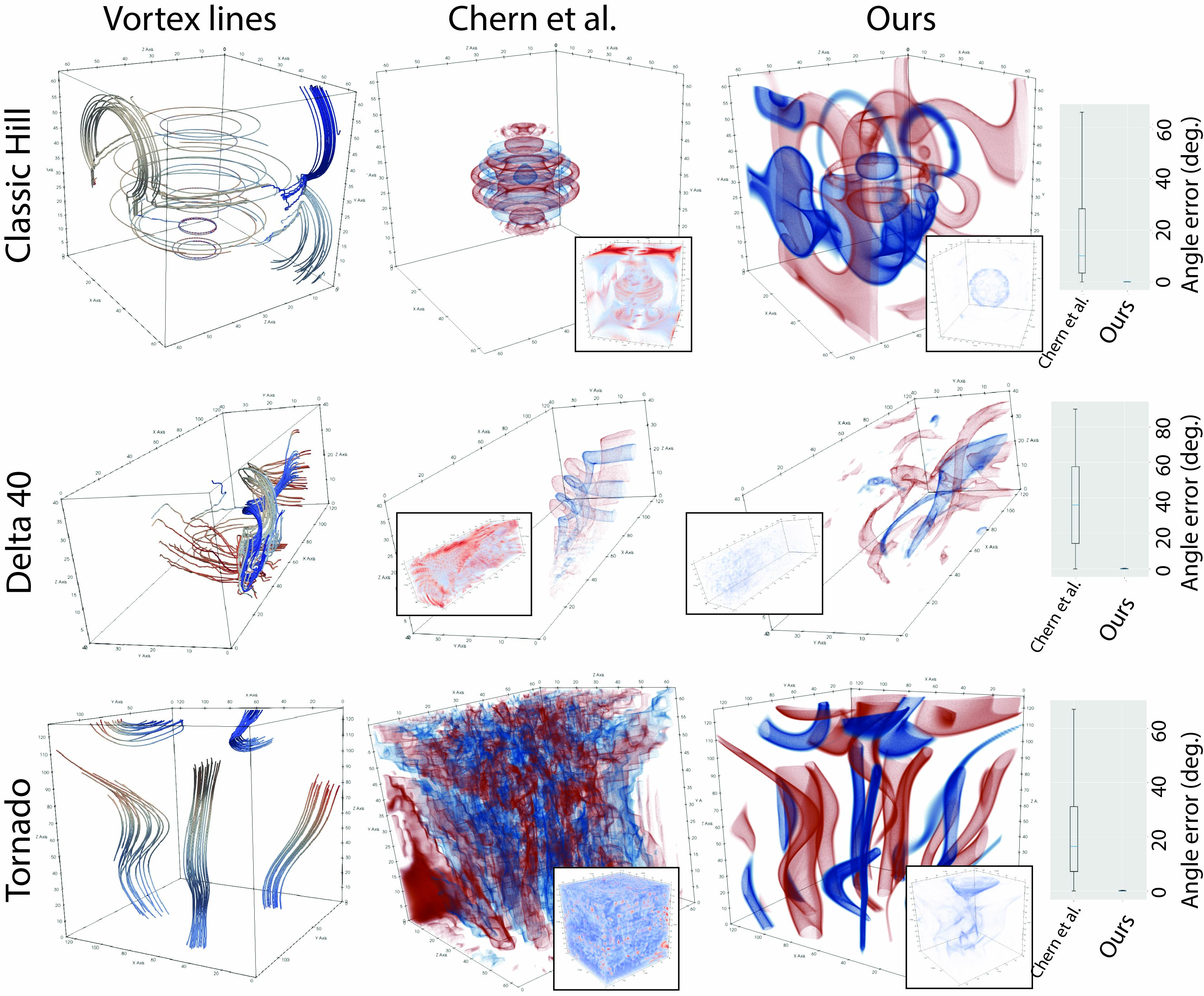}
  \caption{A comparison between representations of vortex tubes from Chern et al. \cite{chern17_insidefluids} and our approach using the $\mathcal{L}_{\perp}$ loss on the vorticity of a vector field. The first column shows vortex lines traced explicitly. Boxplots on the right show the distribution of $Err_{\perp}$ for each method in degrees. Error volume visualizations are inscribed within each render, where blue is less error and red is more error.}
  \label{fig:InsideFluids}
\end{figure}

Shown in \autoref{fig:InsideFluids}, the vortex tubes extracted with the two methods give very different vortex tube visualization.
For the classic hill and delta 40 datasets, the method by Chern et al.\cite{chern17_insidefluids} creates tubes that are very localized to a small part of the domain, whereas our method generates surfaces that span the whole domain.
The quantitative evaluation shows that our approach provides a much smaller median error as well as smaller spread for error across each dataset tested.
In an error-volume analysis, the method by Chern et al.\cite{chern17_insidefluids} has large error outside of the small region of the domain where it creates vortex tubes, whereas our approach has low error throughout, visualized in the error field visualizations inscribed in the renders in \autoref{fig:InsideFluids}.
For example, the method by Chern et al.\cite{chern17_insidefluids} has very low error in the region of the Delta 40 dataset where the extracted vortex tubes are, but has very high error elsewhere.
Lastly, the method by Chern et al. \cite{chern17_insidefluids} fails to provide a useful visualization for vortex tubes of the tornado dataset.

\begin{figure}[ht]
\centering
  \includegraphics[width=\linewidth]{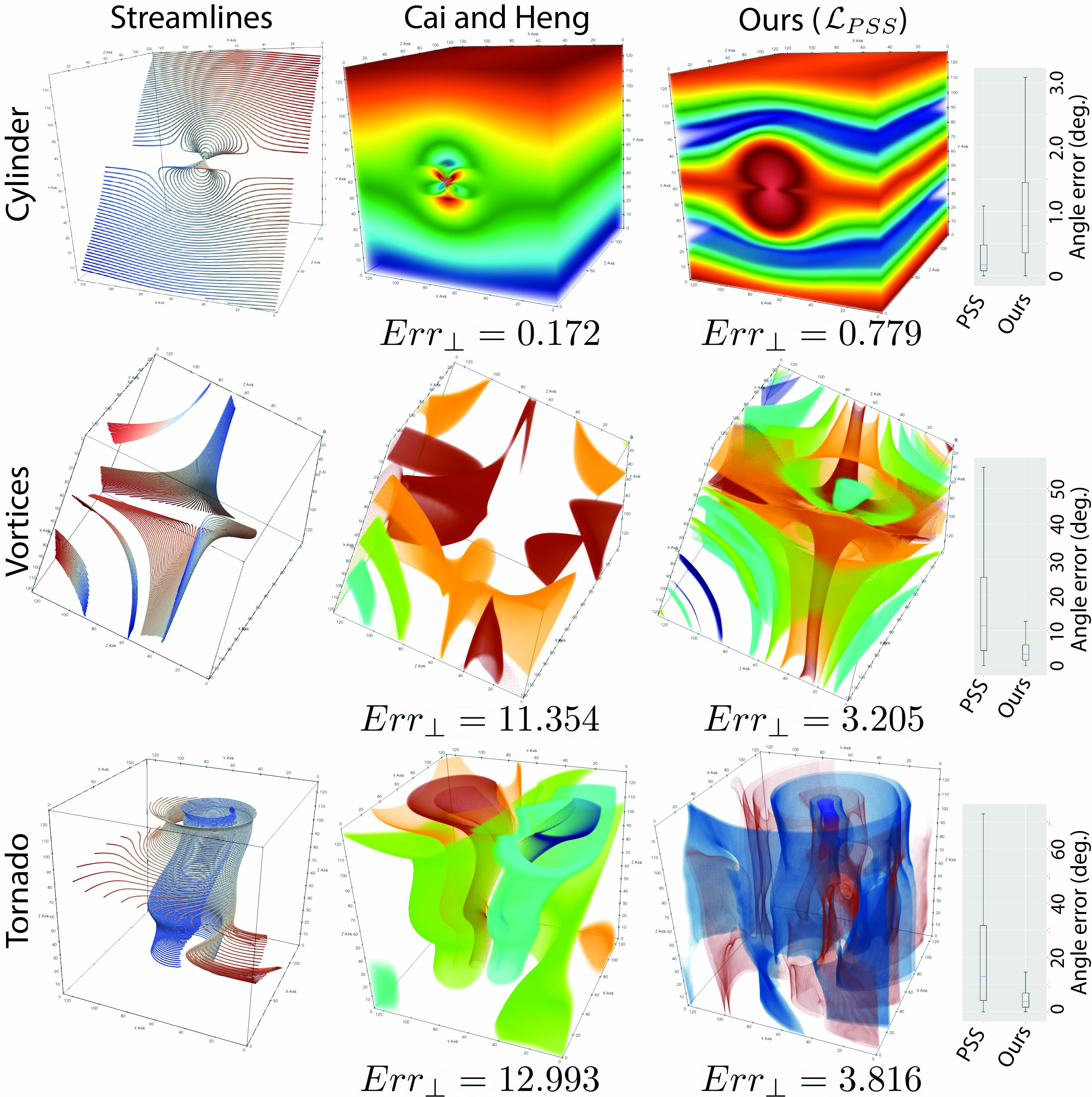}
  \caption{A comparison between Cai and Heng's Principal Stream Surface algorithm \cite{Cai97_principalstreamsurface} and our approach using the $\mathcal{L}_{PSS}$ loss. Listed $Err{\perp}$ represents the median error (in degrees) over all voxels in the volume. Boxplots on the right show the distribution of $Err_{\perp}$ for each method in degrees. Streamlines on the left are colored by integration time.}
  \label{fig:PSS}
\end{figure}

\subsubsection{$\mathcal{L}_{PSS}$ and Principal Stream Surfaces}
\label{PSSevaluation}


A related work by Cai and Heng \cite{Cai97_principalstreamsurface} titled ``Principle Stream Surfaces'' (PSS) finds an implicit solution for stream surfaces in an irrotational field such that the stream surface normals are exactly the normal direction in the Frenet frame. 
Though their solution is restricted to velocity fields that admit a scalar potential for the normal field to a vector field (ours does not have this restriction), we compare their solution with our approach using the $\mathcal{L}_{PSS}$ loss.

Visualizations of the resulting stream functions are shown in \autoref{fig:PSS} with error metrics shown under each volume.
Box plots depicting the error distribution for each method are shown in the rightmost column of \autoref{fig:PSS}, which show that our approach maintains a smaller average error as well as a tighter error spread in all but the cylinder dataset, where the error is relatively small compared with the other datasets, our median being 0.779 degrees compared to PSS 0.172 degrees.
Both algorithms create nearly identical surfaces for the domain outside of the cylinder, which would be in the center of the volume along the z-axis.
For the other two datasets, our solution finds surfaces of interest with lower error than Cai and Heng's method. 
For instance in the vortices dataset, Cai and Heng's method creates the most accurate surfaces in the bottom left of the domain, but fails to capture the correct tube-like structure our method extracts, visible with the red-colored stream surface in the rendering of our approach.
In the tornado dataset, PSS erroneously splits the domain into two sections, whereas our surfaces correctly wind around the domain and converge in the vortex core.

We observe that Cai and Heng's algorithm has lowest error at the origin (index $[0,0,0]$), with error increasing as the scanline algorithm moves away, as expected, whereas the error in our result is clustered closer to higher-frequency features.


\begin{figure}[h]
\centering
  \includegraphics[width=\linewidth]{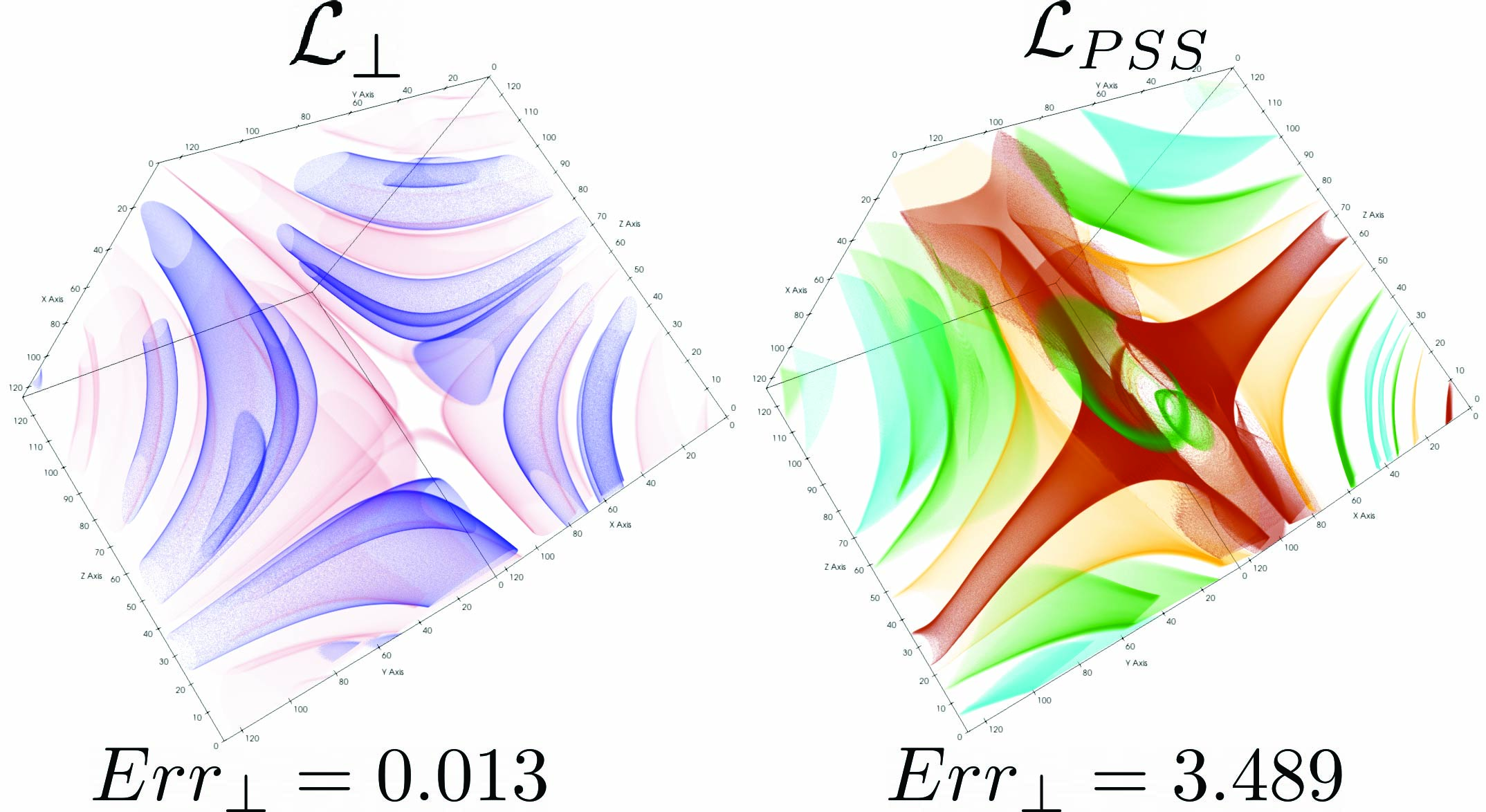}
  \caption{Comparing two stream functions generated for the vortices dataset. The volume on the left was learned using our approach with the $\mathcal{L}_{\perp}$ loss function, while the volume on the right was learned using our approach with $\mathcal{L}_{PSS}$. $Err_{\perp}$ listed is in degrees.}
  \label{fig:lossfunctioncomparison}
\end{figure}

We also visualize in \autoref{fig:lossfunctioncomparison} the difference in stream function learned when using $\mathcal{L}_{PSS}$ as opposed to our non-constrained loss of $\mathcal{L}_{\perp}$ on the same dataset.
Both visualizations in \autoref{fig:lossfunctioncomparison} are stream functions for the vortices dataset, but the left column was trained using $\mathcal{L}_{\perp}$, and the right column was trained using $\mathcal{L}_{PSS}$.
The surfaces created from $\mathcal{L}_{PSS}$ are oriented by the normal direction of the flow, which reveals a different structure than when using $\mathcal{L}_{\perp}$, which creates tube shaped surfaces.
However, more error is introduced by constraining the network to learn a more specific stream function with $\mathcal{L}_{PSS}$.

\begin{figure}[h]
\centering
  \includegraphics[width=\linewidth]{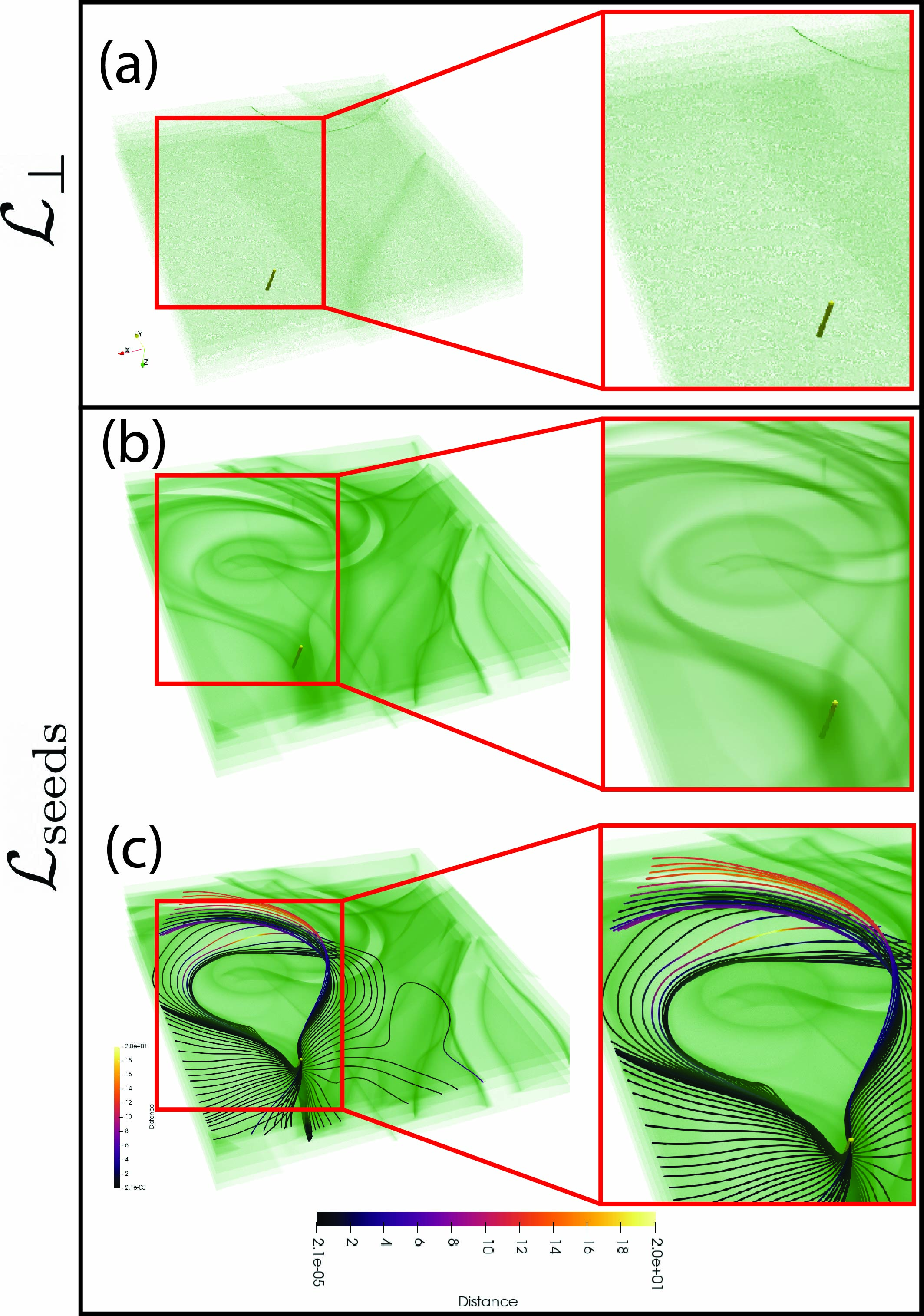}
  \caption{Visualizations of the isosurface at isovalue 0 for the learned stream function from networks trained with $\mathcal{L}_{\perp}$ (a) and $\mathcal{L}_{\text{seeds}}$ (b, c) on the Isabel dataset. 
  (b) and (c) are the same volume rendering with the streamlines disabled or enabled for visual clarity of the stream function. 
  Median distance from the streamlines traced in the original vector field to the stream surface extracted is 0.53 voxels, visualized using the colormap displayed on the bottom.}
  \label{fig:Seeds}
\end{figure}

\subsubsection{Stream Function With Seeding Rake}
\label{seedseval}

Given a seeding rake of interest, our approach can learn a stream function such that an isosurface at isovalue 0 will contain the seeding rake, as described in \autoref{implicitseedingpoints}.
To evaluate our approach, we compare two networks on the same dataset: one trained using $\mathcal{L}_{\perp} +\mathcal{L}_{\text{seeds}}$ and one using $\mathcal{L}_{\perp}$. 
In \autoref{fig:Seeds}, the resulting stream functions are visualized using a spiking transfer function to extract the isosurfaces for isovalue 0.
The network trained with just $\mathcal{L}_{\perp}$ in (a) generates surfaces on the xy-plane with no detail or features around the seeding rake (yellow spheres).
The model trained with the additional $\mathcal{L}_{\text{seeds}}$ term, visualized in (b) and (c), has extracted surfaces that mimic the shape of the streamlines in (c) advected from the seeding rake in the original vector field.
The same volume render is included without streamlines in (b) to reduce occlusion from the streamlines and allow a comparison between the shape of the isosurface and the streamlines in the adjacent image.
We calculate the distance from each point on the streamlines to the closest stream surface point by triangulating the stream surface and calculating the minimum distance from each streamline point to a point on the stream surface.
Visualized in \autoref{fig:Seeds} as the color map on the streamlines, the median distance from the streamlines to the stream surface is 0.53 voxels, while the maximum distance is 19.99 voxels.
Shown by the visualization, the error seems to grow large for some streamlines as they get further from the seeding rake, implying that there may be compounding error for visualized surfaces as the distance from the rake grows.

Although this training routine pushes the network to learn a surface that goes through a seeding rake of interest, the solution stream function can still visualize an infinite number of stream surfaces by varying the isovalue.
We measure the quantitative error of the solution using $Err_{\perp}$ as we do for our other loss functions.
We observe slightly higher error when using $\mathcal{L}_{\text{seeds}}$ as opposed to just $\mathcal{L}_{\perp}$ as shown in \autoref{quanterrortable}. 
However, the error is very small (a maximum of 0.08 degrees in our tested datasets), implying correct stream surface visualization.

\subsubsection{Discussion on Loss Function Choice}
\label{discussion}
\textcolor{black}{Two choices must be made when training a model that determines what stream function will be extracted: (1) is the flow's curvature important to visualize, and (2) is there a specific seeding rake of interest. Question (1) determines if one should use $\mathcal{L}_{\perp}$ or $\mathcal{L}_{PSS}$, and question (2) determines if one should add on $\mathcal{L}_{\text{seeds}}$ to the first loss function.
However, error is another factor that plays a role, as more restrictive loss function choices will result in higher error.}
A comparison of quantitative performance between each loss function is provided in \autoref{quanterrortable}.
$\mathcal{L}_{\perp}$ consistently provides the lowest error stream function compared with the other two choices, but the user has the least control over the extracted stream function.
This accuracy is expected, since $\mathcal{L}_{\perp}$ has the least constraints imposed on the network during training; the network is allowed to learn a gradient which is in any one of the infinite directions orthogonal to the original vector field.

$\mathcal{L}_{PSS}$ may be useful if the stream surfaces extracted when training using $\mathcal{L}_{\perp}$ do not provide enough information about the characteristics of the flow.
Alternatively, two networks trained using $\mathcal{L}_{\perp}$ and $\mathcal{L}_{PSS}$ may be complementary when visualizing stream surfaces.
For instance in our comparison of visualizations provided by the two loss functions in \autoref{fig:lossfunctioncomparison}, both show roughly the same flow features, but in different ways.
Depending on the vector field, one or the other may be more useful, but together \textcolor{black}{they} may improve clarity for flow features.

Finally, $\mathcal{L}_{\text{seeds}}$ may be useful when a seeding rake as well as it's immediate surroundings are of interest.
Isovalue 0 extracts the surface going through the seeding rake, but small changes to the isovalue will generate stream surfaces that may be nearby the seeding rake of interest, allowing quick study of the difference between nearby stream surfaces.
Regardless of loss function choice, the user is completely aware of the accuracy of the result through the calculation of $Err_{\perp}$.

\begin{figure}[ht]
\centering
  \includegraphics[width=\linewidth]{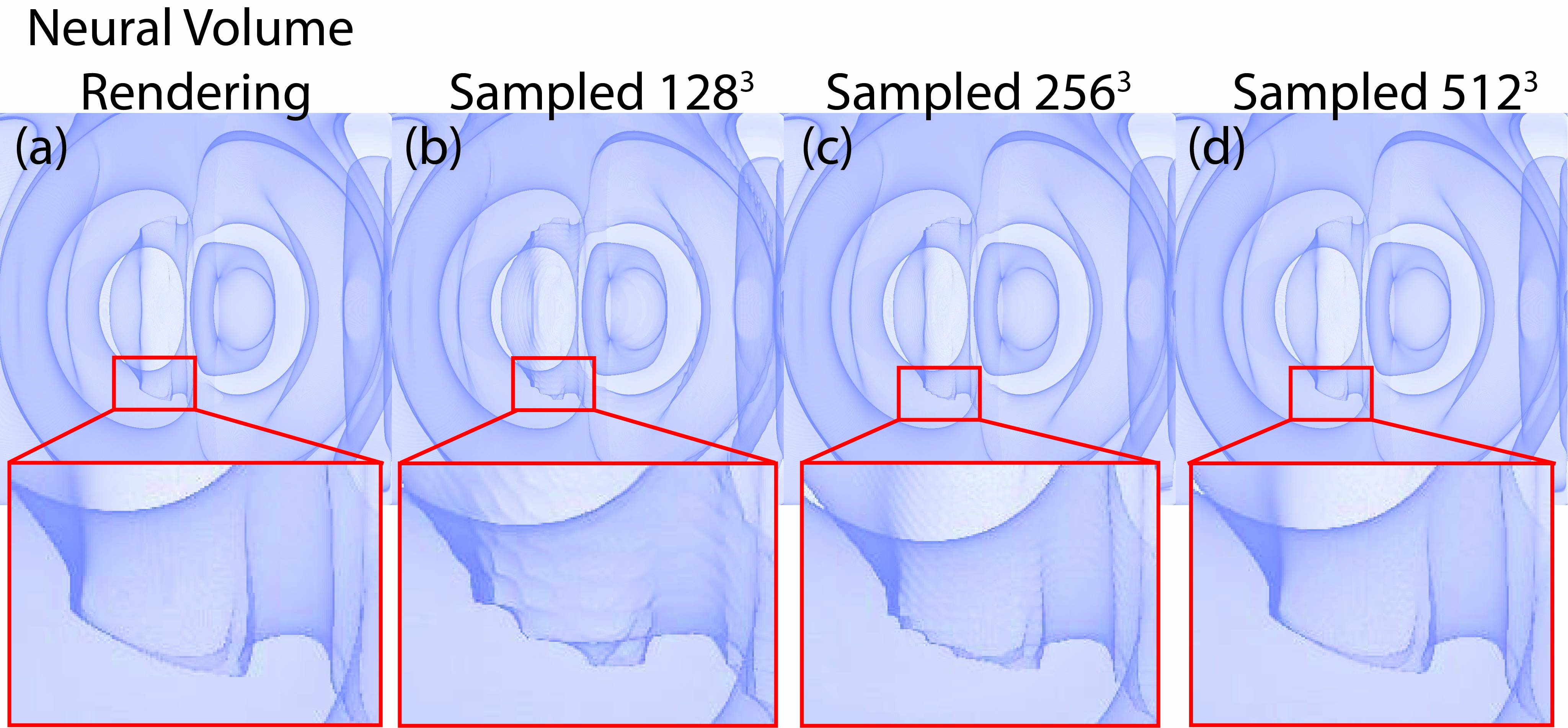}
  \caption{Four renderings of the same pre-trained network that has learned a stream function for the classic hill dataset. Each imaged is rendered at $1024^2$ with $4096$ samples per ray. In (a), the rendering is done by querying the neural network within the volume rendering algorithm, or so-called neural volume rendering. (b)-(d) use the same volume rendering code, but instead of sampling the neural network directly, sample from a pre-evaluated grid of varying sizes, using trilinear interpolation between vertices.}
  \label{fig:neuralrendering}
\end{figure}

\subsection{Visualizing Implicitly Represented Neural Stream Functions}
\label{neuralrendering}

A neural network trained using our approach represents a stream function implicitly as a function $f: \mathbb{R}^3 \rightarrow \mathbb{R}$.
Unlike other approaches that create results for stream functions on a regular grid, our solution is meshless.
Research toward efficiently rendering these meshless INR models has started \cite{lindell21_autoint, sharp22_spelunking, wu22_INRVR}, but we opt to sample the network to a regular-grid for simplicity and ease of use with software such as ParaView.

This presents a new question: what is the best sampling rate to most accurately represent the surfaces of the learned stream function encoded by our INR?
Under-sampling may lead to artifacts on the visualized surfaces, whereas oversampling may be unnecessarily computationally expensive.
To evaluate this, we sample a uniform grid volume from a trained network at 3 sampling rates before traditional volume rendering with trilinear interpolation, and compare the result to directly sampling the neural network during volume rendering (so-called neural volume rendering).
Neural volume rendering will provide the ``ground truth'', since we are directly sampling the network itself during rendering, whereas sampling to a grid and rendering with trilinear interpolation will only estimate the underlying value the network would create.

In our test, we use our trained neural network that learned the stream function for the classic hill dataset, which has a resolution of $64^3$.
From this trained network, we uniformly sample three discrete volumes of sizes $128^3$, $256^3$, and $512^3$.
\autoref{fig:neuralrendering} shows the result of rendering these three volumes compared with directly sampling the neural network during volume rendering.
Each image is rendered at a resolution of $1024^2$ with a ray depth of $4096$ uniformly spaced samples per ray.
The results show that even though the original volume was only $64^3$, the volumes sampled at $128^3$ and $256^3$ introduce artifacts on the surface, and in order to properly visualize the result from the neural network with traditional volume rendering on a discrete grid, a grid size of $512^3$ is necessary.

\section{Conclusion, Limitations, and Future Work}

We present an improvement over state of the art implicit stream surface extraction methods. 
Our neural network method learns an unknown stream function given a vector field, from which isosurfaces extract stream surfaces.
We also offer training routines that will create stream functions with flow curvature-aware surfaces or stream functions with surfaces going through a seeding rake.
Our evaluation shows that our results have lower error than other implicit methods, and we provide insight into the sampling rate necessary to visualize the neural network after training.

Our approach has a few limitations.
Though the error of the principal stream function solutions trained using $\mathcal{L}_{PSS}$ are still acceptable, the error is significantly larger than the error of the unconstrained models trained using $\mathcal{L}_{\perp}$, which presents a trade-off between extracted stream surface expressiveness and accuracy.
The loss function $\mathcal{L}_{PSS}$ is inherently more challenging to learn than $\mathcal{L}_{\perp}$ since $\mathcal{L}_{PSS}$ looks for a gradient which is exactly parallel to the principal normal direction, whereas there are infinite orthogonal directions to the vector field that minimize $\mathcal{L}_{\perp}$.
As neural network architectures and training routines improve, we expect the error of our method's results to decrease, but the principal stream function error will likely remain relatively higher.
Similarly, another limitation is the larger error when training using a seeding rake.
As with the principal stream function training, adding this regularizing term $\mathcal{L}_{\text{seeds}}$ increases the difficulty for the network to learn a solution, and so error will increase.
\textcolor{black}{Additionally, as the data become larger-scale, the neural network size necessary to adequately learn a stream function must also increase, along with batch size, meaning longer training times.
As an option, one may crop the data to a smaller subset for higher quality stream function extraction without sacrificing training time.}
Another drawback is the issue of occlusion when visualizing the learned stream function. 
Stream function values tend to repeat in different spatial regions, causing one isovalue to create many stream surfaces, making it challenging to visualize a single stream surface.
\textcolor{black}{Lastly, our method seems to capture global shape well, but may not generate accurate surfaces near complex critical points like sources and sinks.}

In the future, implicit neural representations that train within seconds, such as those used by Müller et al. \cite{mueller22_instant}, can accelerate stream function solution time as well as enable interactive neural rendering at framerates, which are currently two limitations of our solution using SIREN.
A combination of Müller et al.'s method \cite{mueller22_instant} with the training strategy presented in AutoInt \cite{lindell21_autoint} may further improve training speed for our network.
\textcolor{black}{With a modified loss function to $\mathcal{L}_{\perp}$ that maximizes the magnitude of the inner product instead of minimizes, our approach could be extended to generate as-perpendicular-as-possible surfaces like the method by Schulze et al. \cite{APAP_schulze12}.}
Another future research direction is implicitly representing other surfaces such as time-surfaces, streak-surfaces, and path-surfaces as level sets of the scalar field solution learned by an implicit neural representation.

\acknowledgments{


This work is supported in part by the US Department of Energy SciDAC program DE-SC0021360, National Science Foundation Division of Information and Intelligent Systems IIS-1955764, and National Science Foundation Office of Advanced Cyberinfrastructure OAC-2112606. 
This work is also supported by Advanced Scientific Computing Research, Office of Science, U.S. Department of Energy, under Contract DE-AC02-06CH11357, program manager Margaret Lentz.
}

\bibliographystyle{abbrv-doi}

\bibliography{template}
\end{document}